# Practical Evaluation of Wize and Bluetooth 5 Assisted RFID for an Opportunistic Vehicular Scenario

1st Ángel Niebla-Montero
*Department of Computer Engineering*
*Faculty of Computer Science, Universidade da Coruña*
A Coruña, Spain
0000-0002-1648-9440

2nd Iván Froiz-Míguez
*Department of Computer Engineering*
*Faculty of Computer Science, Universidade da Coruña*
A Coruña, Spain
0000-0003-4068-8842

3rd Paula Fraga-Lamas
*Department of Computer Engineering*
*Faculty of Computer Science, Universidade da Coruña*
*Centro de investigación CITIC, Universidade da Coruña*
A Coruña, Spain
0000-0002-4991-6808

4th Tiago M. Fernández-Caramés
*Department of Computer Engineering*
*Faculty of Computer Science, Universidade da Coruña*
*Centro de investigación CITIC, Universidade da Coruña*
A Coruña, Spain
0000-0003-2179-5917

*Abstract*—Wireless communications are critical in the constantly changing environment of IoT and RFID technologies, where thousands of devices can be deployed across a wide range of scenarios. Whether connecting to cloud servers or local fog/edge devices, maintaining seamless communications is difficult, especially in demanding contexts like industrial warehouses or remote rural areas. Opportunistic networks, when combined with edge devices, provide a possible solution to this challenge. These networks enable IoT devices, particularly mobile devices, to redirect information as it passes via other devices until it reaches an edge node. Using different communication protocols, this paper investigates their effects on response times and total messages received for a opportunistic assisted RFID system. Specifically, this article compares two communications technologies (Bluetooth 5 and Wize) when used for building a novel Opportunistic Edge Computing (OEC) identification system based on low-cost Single-Board Computers (SBCs). For such a comparison, measurements have been performed for quantifying packet loss and latency. The tests consisted in two experiments under identical conditions and scenarios, with a node located roadside, transmitting identification information, and a node located inside a moving vehicle that was driven at varying vehicle speeds. The obtained results show for Bluetooth 5 average latencies ranging between 700 and 950 ms with packet losses between 7% and 27%, whereas for Wize the average delay as between 150 and 370 ms with packet losses between 20% and 52%.

I. INTRODUCTION

Many of the IoT and RFID devices deployed throughout the world have limited storage, processing capacity and energy consumption, forcing them to rely on distant devices for intense computational operations or additional security features [1]. Moreover, although smart IoT devices need to communicate with each other, wireless connection coverage is not always accessible. For instance, the intricate nature of highly metallic industrial environments usually disrupts wireless communications [2]. To solve these difficulties, this paper proposes an Opportunistic Edge Computing (OEC) identification architecture that allows IoT devices to collaborate and share resources and services. OEC systems can identify deployed RFID, IoT or Industrial IoT (IIoT) devices and deliver Edge Computing services as needed [3].

Among the different types of RFID, are those assisted by other technologies, such as Bluetooth, Wi-Fi, Zigbee or Wize. Thus, assisted RFID provides a way to leverage the strengths of both kinds of technologies. This paper focuses on two assisted-RFID systems for vehicular environments: one Bluetooth-based and another one Wize-based. The goal is to use these two base technologies to communicate data to nearby devices, resulting in long ranges and continuous monitoring. As a result, the purpose of this paper is to compare the latency and packet loss of two opportunistic systems: a Bluetooth 5-assisted [4] and a Wize-assisted [5] RFID system. In such systems, roadside devices transmit periodically information to the circulating vehicles, which can use the received data for identifying the transmitting devices, as well as for collecting other relevant data in an opportunistic way.

II. STATE OF THE ART: OPPORTUNISTIC MOBILE EDGE COMPUTING SYSTEMS

OEC systems have been already deployed for implementing various applications, such as identification and monitoring. For instance, in [6] the authors use an OEC system for

This work has been funded by grant PID2020-118857RA-100 (ORBALLO) funded by MCIN/AEI/10.13039/501100011033.

wildlife monitoring. In such paper the authors combine Lora and Bluetooth to achieve ultra-low power consumption while enabling real-time wildlife monitoring. This is accomplished by processing data locally on the devices (via BLE) before transmitting only critical information across the long-distance LoRa network to identify the assets to be monitored, similar to an assisted RFID system. This reduces the amount of data that must be transmitted, resulting in longer battery life for wildlife monitoring devices.

There are also examples of OEC systems for smart cities [7]. Such a paper investigates Vehicular Fog Computing (VFC) in the context of smart cities. VFC uses vehicle processing power to handle tasks that are offloaded from mobile devices. This approach uses Fog Computing principles to bring computation closer to users (vehicles), resulting in faster response times than traditional cloud computing. The paper focuses on challenges such as short-lived or intermittent connections between vehicles and users, as well as time-consuming multi-hop forwarding between vehicles that can result in packet loss.

In addition, there are theoretical articles on IoT computing architectures. For instance, two papers investigate alternative computing architectures to overcome the limitations of Cloud Computing for specific IoT applications: in [8] it is proposed a more adaptable Fog Computing system, while in [9] the authors demonstrate its advantages in optimizing energy usage for smart agriculture. Specifically, in [8] it is proposed a Fog Computing system that can function even when the capabilities of the collaborating devices vary (like it occurs in a typical opportunistic scenario). The authors also propose a network architecture that relies on virtual clusters and peer-to-peer connections to simplify the management of complex underlying networks. Simulations in their paper assess the feasibility of constructing such an opportunistic fog computing network. On the other hand, in [9] the authors explore an architecture that combines Edge, Fog and Cloud Computing to improve energy efficiency in smart agriculture. Sensors collect data for real-time processing in agricultural operations. Their architecture uses Edge and Fog layers for initial processing, which reduces the workload on the Cloud and saves energy. The study suggests a significant reduction in energy consumption and carbon emissions when compared to traditional methods.

Other articles focus on specific features, such as [10], which presents a novel approach for managing the growing demands of the IoT by combining blockchain technology, distributed cloud architecture, fog nodes and Software-Defined Networks (SDNs). The results of the research show that, when compared to traditional cloud computing infrastructure, the proposed model is more efficient at reducing end-to-end delay between devices and offloading data to the cloud.

III. DESIGN AND IMPLEMENTATION OF THE SYSTEM

A. Communications architecture

Figure 1 shows the proposed OEC IoT architecture, which is divided into three layers:
- IoT Network Layer: this layer consists of several IoT or RFID networks that share data with higher levels as needed. Nodes equipped with sensors and actuators monitor and interact with their surroundings. Communications with the upper layers is opportunistic, requiring devices to determine when upper layer devices are accessible for service access.
- OEC Smart Gateway Layer: Made up of Fog Computing gateways, this layer opportunistically offers services to the IoT Network Layer with lower latency than distant cloud services. Data from IoT devices is saved on a network of Distributed Hash Tables (DHTs) shared by gateways.
- Cloud Layer: Offering services that go beyond what OEC smart gateways can do, such as compute-intensive processing or large-scale data storage.

The OEC Smart Gateway Layer provides critical services such as peer discovery, peer routing and data routing. Peer discovery enables the finding of peer addresses using information supplied by other peers. Peer routing provides an interface to determine the peers to which messages should be routed. When the receiving peer is out of range of the sender peer, data routing methods forward messages to them. These functions are provided through libp2p [11]. Libp2p was chosen for this critical role due to its versatility and transport-agnostic nature. It provides a comprehensive set of protocols for developing peer-to-peer applications and is modular, allowing its easy incorporation into OEC architectures. In addition, Libp2p offers solutions for packet transport, security, peer-to-peer routing and content discovery, making it ideal for the creation of robust and flexible OEC systems.

Regarding the cloud layer, it provides routing services, enabling the communication across various IoT and RFID networks via scattered gateways. This enables to provide seamless connectivity and communications across multiple distant IoT/RFID networks.

For the IoT network layer of this paper, two communications technologies were selected: Bluetooth 5 and Wize. Such technologies are later evaluated through experiments to determine which one is better in a specific scenario: when sending identification and relevant information from a roadside beacon to a moving vehicle. Table I shows a comparison of the main characteristics of each of the protocols chosen for the experiments.

B. Bluetooth 5-assisted RFID hardware

A distributed communications system allows for avoiding part of the limitations of Cloud Computing (e.g., dependency on a point of failure, potential communications bottleneck, sometimes difficult scalability), since they eliminate the need for a central connection to the cloud. Several wireless protocols enable distributed communications, notably those that operate in the unrestricted and frequently utilized 2.4 GHz Industrial Scientific Medical (ISM) band. Such protocols provide distributed communications, which overcome short-range restrictions and support a high number of nodes. For this paper, Bluetooth 5 LE is used because it is one of the most energy-efficient wireless communications technologies [12], with one of the highest transmission/energy ratios. Moreover, Bluetooth 5 LE can be completely decentralized, allowing nodes to communicate via advertisements through BLE Mesh, making it ideal for opportunistic communications.

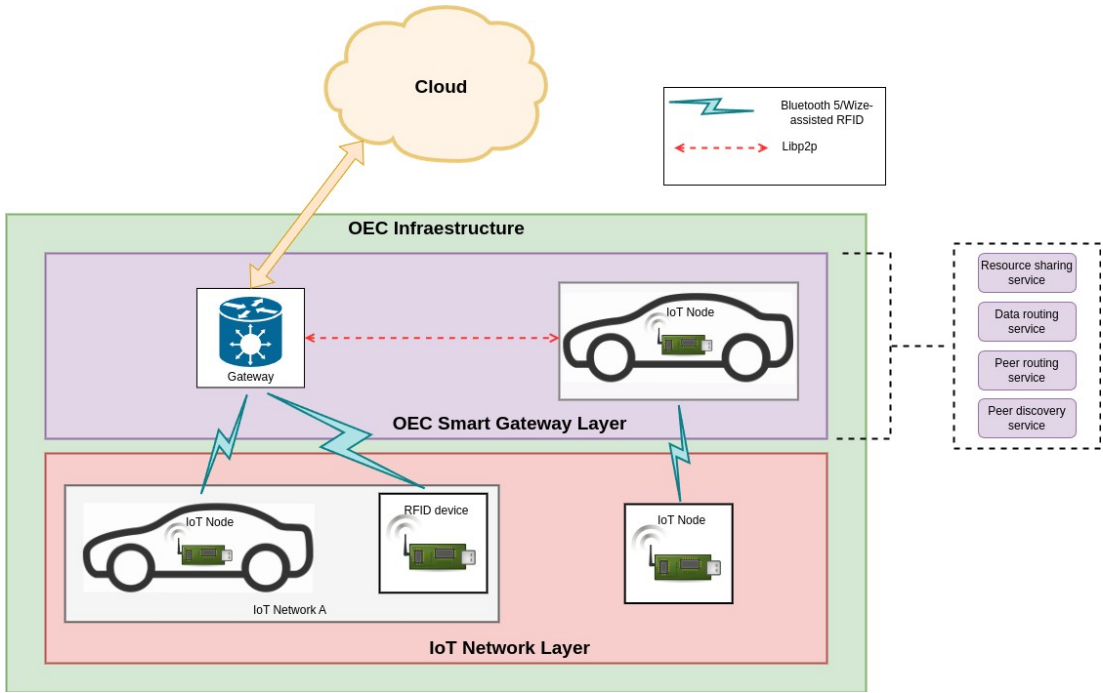

Fig. 1. Designed OEC communications architecture.

The use of Bluetooth 5 with BLE Mesh can enhance communications and enable new applications. For example, Bluetooth 5 significantly improves the features of previous 4.x versions, increasing bandwidth, introducing extended advertisements and range, and offering a long-range mode that increases the sensitivity and range of communication. This functionality is particularly useful for opportunistic systems that run over long distances or in electro-magnetic "noisy" situations, such as industrial scenarios.

### C. Wize-assisted RFID hardware

In contrast to Bluetooth, Wize is a communications protocol that has only been on the market for a short time. The Wize protocol is the result of a new LPWAN (Low Power Wide Area Network) solution built on Wireless M-Bus at 169 MHz. This technology is intended to be used in Smart Cities and industrial IoT applications in general.

Lower frequencies, such as 169 MHz, are advantageous for wireless communications due to their longer range and better penetration through obstacles such as buildings, being especially useful in situations where line-of-sight communications are not possible. On the downside, Wize has low data rates, longer transmission durations (several hundred milliseconds), and high transmission power (500 mW). Despite this, it remains as a low-cost option for sporadic transmissions.

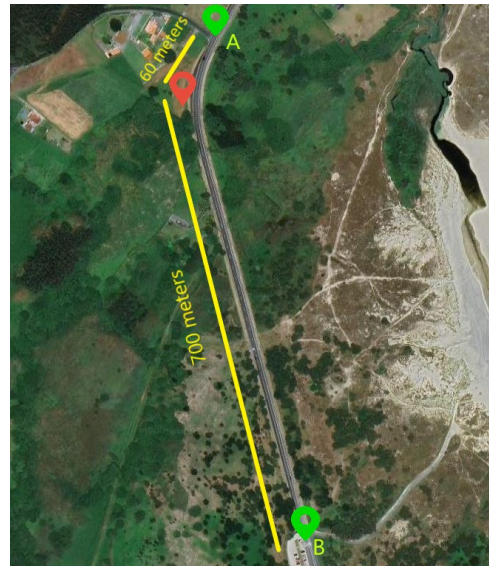

Fig. 2. Testing scenario.

## IV. EXPERIMENTS

To compare the performance of the proposed systems, two experimental testbeds were created. Such testbeds are able to measure the latency and packet losses of the created OEC system by transmitting identification information from a RSU (Roadside Unit) placed nearby the road (indicated with a red pin in Figure 2) to a node carried by a moving car. The car was driven at three vehicular speeds (30, 50 and 70 km/h, which are roughly 18.6, 31 and 43.5 mph) between the positions A and B indicated in Figure 2 (the itinerary started at point A, reached point B, and then returned to point A).

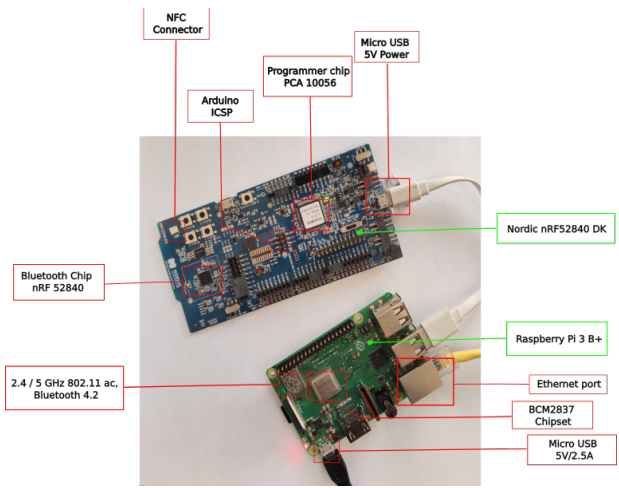

Fig. 3. Components of the Bluetooth 5 testbed IoT node.

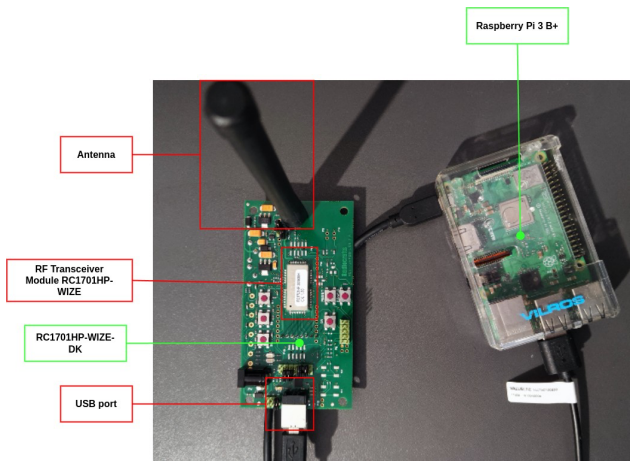

Fig. 4. Components of the Wize testbed IoT node.

TABLE I
FEATURE COMPARISON BETWEEN BLUETOOTH 5 AND WIZE PROTOCOLS.

| Aspect | **Bluetooth 5 (BLE)** | **Wize** |
| --- | --- | --- |
| Frequency Band | 2.4 Ghz | 169 Mhz |
| Range | Middle range (up to 400 meters) | Long range (up to several kilometers) |
| Interference | From other devices in the same band, such as Wi-Fi and microwaves | Narrowband radio improves the selectivity, reducing interference and blocking |
| Data Rate | Between 125 Kbps and 2 Mbps | Between 2.4 and 6.4 Kbps |
| Power Consumption in transmission | 20 mA | 400 mA |
| Compatibility | High compatibility | Fewer compatible devices |
| Specific Applications | IoT, IIoT, consumer wearables, remote monitoring | Smart city wireless networks, fluids and gas metering, heavy industry |

*A. Bluetooth 5-assisted RFID experimental setup*

This testbed consisted of an IoT node (shown in Figure 3) built on a SBC (Raspberry Pi 3B+) and a Nordic nRF52840 development kit that provided Bluetooth Mesh capabilities via a serial connection. The node was modified with the developed software, which was then integrated into the Nordic nRF Mesh app for Android. In addition, a RSU was built with a Raspberry Pi 3B. The RSU, like the other OEC's IoT nodes, enabled Bluetooth Mesh communication via two Nordic nRF52840 development kits. Both nodes were set in the same mesh group previously to the beginning of the experiments.

The RSU periodically transmitted packages containing the node's identification information to the mobile IoT node via Bluetooth 5 (BLE).

*B. Wize-assisted RFID experimental setup*

For the Wize-based experiments, the mobile node (show in Figure 4) and the RSU (shown in Figure 5), consisted of two SBCs (Raspberry Pi 3B and Raspberry Pi 3B+, respectively) and two Wize modules (RC1701HP-Wize-DK [13]). The configuration of the Wize modules was the same for all the performed tests (Operating frequency: 169.431250 MHz and Transmit power: +27 dBm), except for the data rate, which was modified to determine how its modification impacted the communications. In particular, two data rates were evaluated: 2.4 kbps (GFSK) and 6.4 kbps (4GFSK).

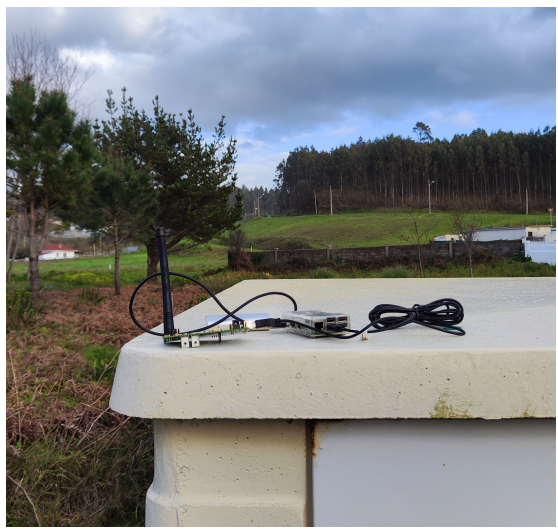

Fig. 5. RSU in the test scenario.

## C. Packet loss rate

Figure 6 shows that, when delivering data over Bluetooth 5, 11% of packets were lost at 30 km/h, 27% at 50 km/h, and 7% at 70 km/h.

Figure 7 shows that, when using Wize with a data rate of 2.4 kbps, 20% of the packets were lost at 30 km/h, 20% at 50 km/h, and 43% at 70 km/h. With a data rate of 6.4 kbps, 42% of the packets were not received at 30 km/h, 50% at 50 km/h and 52% at 70 km/h.

The results of the Bluetooth 5 tests were quite satisfactory, except for a speed of 50 km/h, where the percentage of lost packets was really high in comparison to the rest of the tests. Such a behavior was attributed to a temporary disruption of communications during testing, but it has been preserved in the results to reflect what actually can happen in a real scenario. On the other hand, the Wize tests provided worst results when transmitting a data rate of 6.4 kbps, with more than half of the packets lost in all cases (compared to an average of 27% for 2.4 kbps). In comparison, the average of lost packets in the experiment with Bluetooth 5 was 15%, well below the best results for Wize.

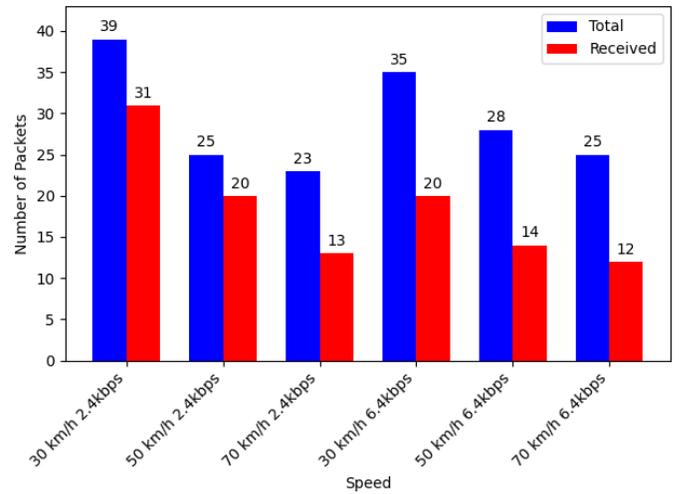

Fig. 7. Total packets sent versus packets received with Wize.

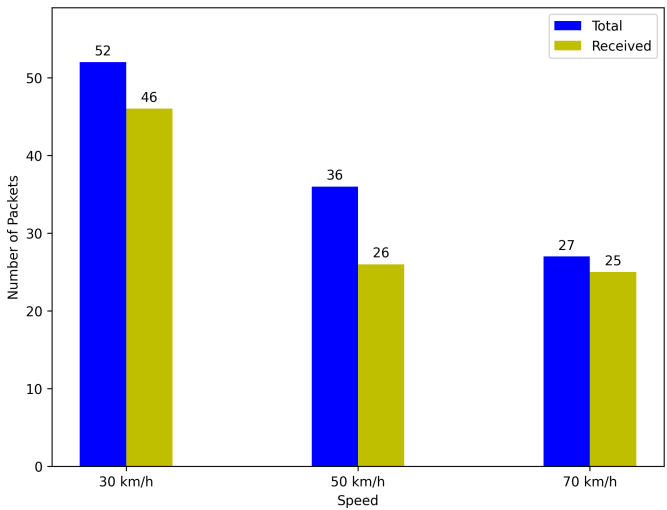

Fig. 6. Total packets sent versus packets received with Bluetooth 5.

## D. Transmission latency

Regarding the latency of the received packets, Figures 8 and 9 show the latency for each of the Bluetooth 5 and Wize experiments respectively. Note that, in such Figures, the number of plotted latencies is different due to the vehicle speed: the lower the speed, the higher the number of packets that was transmitted/received, since the time dedicated to moving through the test road was longer.

Figure 8 shows for the Bluetooth 5 experiment that the obtained delay was in the 700-900 ms range on average. The minimum average delay for 30 km/h was 716 ms, while the maximum was 955 ms (for 70 km/h).

Figure 9 depicts the latency for the Wize experiment. Such a Figure shows two groups that can be distinguished by their data rate. For 2.4 kbps, the obtained average latency was about 370 ms, whereas for 6.4 kbps such an average latency was 150 ms. Therefore, as expected, the higher the data rate, the shorter the latency.

Therefore, the results presented in Figures 8 and 9 show that Wize obtains a very low latency in comparison to the one obtained by Bluetooth 5 in the selected scenario. Specifically, the average latency obtained by Bluetooth 5 is more than 5 times higher than the one obtained for Wize.

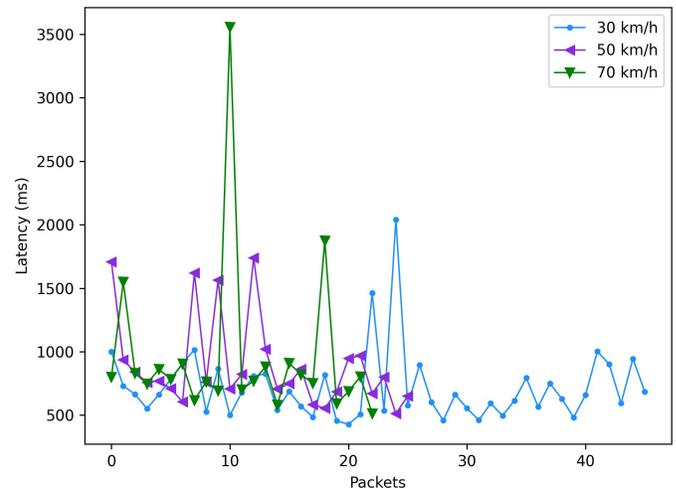

Fig. 8. End-to-end latency at different speeds with Bluetooth 5.

## V. FUTURE WORK

The preliminary research analysis performed in this paper will be extended as follows:
- A complete review will be performed on the state-of-the art of similar system to compare OEC and non-OEC systems.

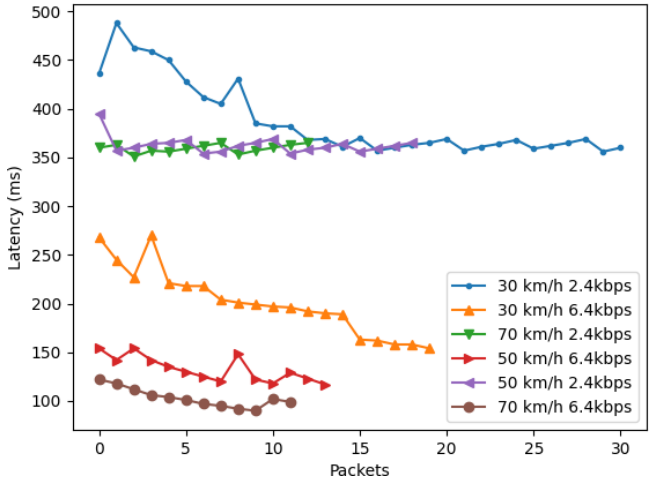

Fig. 9. End-to-end latency at different speeds with Wize.

- A thorough comparative analysis will be carried out on the consumption of the developed assisted-RFID systems.
- Additional studies will be performed in different scenarios (e.g., in hostile industrial locations where communications interference is common, like it occurs in a shipyard).
- Methods for reducing latency and packet loss will be investigated.

## VI. Conclusions

This paper described an innovative OEC system based on mobile nodes that has been specifically designed for mobile IoT applications. After describing the communications architecture, test results have been provided that show the system's practicality. Specifically, the performed experiments investigated the transmission of identification data in a vehicular scenario by using Bluetooth 5 and Wize at different speeds. The results revealed a trade-off between packet loss and latency. The Bluetooth 5-assisted RFID system offered lower average packet loss, particularly at slower speeds, but suffered from significantly higher latency. In contrast, the Wize-assisted RFID system provided much lower latency, especially at a higher data rate, but came at the cost of increased packet loss. Therefore, the optimal choice between these technologies depends on the application's specific needs. If reliable data transmission is critical, Bluetooth 5 could be preferable, especially at slower speeds. However, if low latency is crucial, Wize could be a better option despite the potential for higher packet loss. Furthermore, it should be noted that the Wize system consumes 20 times more energy than Bluetooth 5 in transmission, implying that it is only suitable for IoT nodes with access to the grid or to an adequate charging system.

## References

[1] P. Fraga-Lamas and T. M. Fernández-Caramés, "Reverse engineering the communications protocol of an RFID public transportation card," 2017 IEEE International Conference on RFID (RFID), Phoenix, AZ, USA, 2017, pp. 30-35.
[2] P. Fraga-Lamas, T. M. Fernández-Caramés, D. Noceda-Davila and M. Vilar-Montesinos, "RSS stabilization techniques for a real-time passive UHF RFID pipe monitoring system for smart shipyards," 2017 IEEE International Conference on RFID (RFID), Phoenix, AZ, USA, 2017, pp. 161-166.
[3] Á. Niebla-Montero, I. Froiz-Míguez, P. Fraga-Lamas, and T.M. Fernández-Caramés, "Practical Latency Analysis of a Bluetooth 5 Decentralized IoT Opportunistic Edge Computing System for Low-Cost SBCs," Sensors, vol. 22, 8360, 2022.
[4] Bluetooth Core Specification Version 5.0 (2021, March). [Online]. Available: https://www.bluetooth.com/bluetooth-resources/bluetooth-5-go-faster-go-further/.
[5] The Wize Protocol, LPWAN for Smart Cities (2018, March). [Online]. Available: https://radiocrafts.com/uploads/WP016 Wize \_ Protocol For LPWAN.pdf.
[6] E. D. Ayele, N. Meratnia and P. J. M. Havinga, "Towards a New Opportunistic IoT Network Architecture for Wildlife Monitoring System," 9th IFIP International Conference on New Technologies, Mobility and Security (NTMS). Paris, France, pp. 1–5, February 2018.
[7] C. Tang, X. Wei, C. Zhu, Y. Wang and W. Jia, "Mobile Vehicles as Fog Nodes for Latency Optimization in Smart Cities," IEEE Transl. V. Tech. vol. 69, pp. 9364–9375, Sept 2020.
[8] R. Silva, J. S. Silva and F. Boavida, "Opportunistic fog computing: Feasibility assessment and architectural proposal," IFIP/IEEE Symposium on Integrated Network and Service Management (IM). Portugal, pp. 510–516, May 2017.
[9] H. A. Alharbi and M. Aldossary, "Energy-Efficient Edge-Fog-Cloud Architecture for IoT-Based Smart Agriculture Environment," IEEE Access, vol. 9, pp. 110480–110492, July 2021.
[10] P. K. Sharma, M. -Y. Chen and J. H. Park, "A Software Defined Fog Node Based Distributed Blockchain Cloud Architecture for IoT," IEEE Access, vol. 6, pp. 115–124, September 2018.
[11] B. Guidi, A. Michienzi, and Ricci, L. "A libP2P Implementation of the Bitcoin Block Exchange Protocol," [2nd International Workshop on Dist. Inf. Com. Good , New York, USA, 1–4, 2021]
[12] M. Siekkinen, M. Hiienkari, J. K. Nurminen and J. Nieminen, "How low energy is bluetooth low energy? Comparative measurements with ZigBee/802.15.4," IEEE Wir. Comm. Net. Conf. Work. (WCNCW). France, pp. 232–237, April 2012.
[13] WIZE High Power RF Transceiver Module (2022, March) . [Online]. Available: https://radiocrafts.com/uploads/RC1701HP-WIZE_Datasheet.pdf.